\PassOptionsToPackage{unicode}{hyperref}
\PassOptionsToPackage{hyphens}{url}
\documentclass[
  11pt,
]{article}
\usepackage{amsmath,amssymb}
\usepackage{iftex}
\ifPDFTeX
  \usepackage[T1]{fontenc}
  \usepackage[utf8]{inputenc}
  \usepackage{textcomp} 
\else 
  \usepackage{unicode-math} 
  \defaultfontfeatures{Scale=MatchLowercase}
  \defaultfontfeatures[\rmfamily]{Ligatures=TeX,Scale=1}
\fi
\ifPDFTeX\else
\fi
\IfFileExists{upquote.sty}{\usepackage{upquote}}{}
\IfFileExists{microtype.sty}{
  \usepackage[]{microtype}
  \UseMicrotypeSet[protrusion]{basicmath} 
}{}
\usepackage{xcolor}
\usepackage{longtable,booktabs,array}
\usepackage{calc} 
\usepackage{etoolbox}
\makeatletter
\patchcmd\longtable{\par}{\if@noskipsec\mbox{}\fi\par}{}{}
\makeatother
\IfFileExists{footnotehyper.sty}{\usepackage{footnotehyper}}{\usepackage{footnote}}
\makesavenoteenv{longtable}
\usepackage{graphicx}
\usepackage{pos}
\usepackage[absolute,overlay]{textpos}

\author[a]{Laurence Sebastian Bowes}
\author[a]{Vincent Drach}
\author[b]{Patrick Fritzsch}
\author*[c]{Sofie Martins} 
\author[c]{Antonio~Rago}
\author[d, e]{Fernando Romero-López}

\affiliation[a]{Centre for Mathematical Sciences, University of Plymouth, England, UK}
\affiliation[b]{School of Mathematics, Trinity College Dublin, Ireland}
\affiliation[c]{$\hbar$QTC \& IMADA, University of Southern Denmark, Campusvej 55, 5230 Odense M, Denmark}
\affiliation[d]{Center for Theoretical Physics, Massachusetts Institute of Technology, USA}
\affiliation[e]{Albert Einstein Center, Institute for Theoretical Physics, University of Bern, 3012 Bern, Switzerland}

\emailAdd{laurence.bowes@plymouth.ac.uk}
\emailAdd{vincent.drach@plymouth.ac.uk}
\emailAdd{fritzscp@tcd.ie}
\emailAdd{martinss@imada.sdu.dk}
\emailAdd{rago@qtc.sdu.dk}
\emailAdd{fernando.romero-lopez@unibe.ch}

\abstract{The $SU(2)$ gauge group with two fundamental flavors is a candidate for a composite Higgs extension of the Standard Model. Central to Higgs phenomenology is a non-perturbative determination of observables of the theory, such as the decay constant of the pseudo-Nambu-Goldstone Bosons. We present preliminary results for the continuum limit of the pseudoscalar decay constant using a mixed-action setup, with non-perturbatively improved stabilized Wilson Fermions on the sea, and maximally twisted valence quarks. Pivotal to this study is the recent porting of our simulation suite HiRep to GPU architecture.

\begin{textblock}{20}(15.0,1.70)
MIT-CTP/5765
\end{textblock}%

}

\FullConference{The 41st International Symposium on Lattice Field Theory (LATTICE2024)\\
 28 July - 3 August 2024\\
Liverpool, UK\\}

\usepackage{flafter}
\usepackage{float}
\usepackage{pos}
\usepackage{natbib}
\usepackage{subfigure}
\usepackage{booktabs}
\usepackage{siunitx}
\usepackage{longtable}
\usepackage{array}
\usepackage{multirow}
\usepackage{wrapfig}
\usepackage{colortbl}
\usepackage{pdflscape}
\usepackage{tabu}
\usepackage{threeparttable}
\usepackage{threeparttablex}
\usepackage[normalem]{ulem}
\usepackage{makecell}
\usepackage{xcolor}
\ifLuaTeX
  \usepackage{selnolig}  
\fi
\usepackage[]{natbib}
\nocite{*}
\usepackage{bookmark}
\IfFileExists{xurl.sty}{\usepackage{xurl}}{} 
\hypersetup{
  pdftitle={Determination of the pseudoscalar decay constant from $SU(2)$ with two fundamental flavors},
  hidelinks,
  pdfcreator={LaTeX via pandoc}}

\title{Determination of the pseudoscalar decay constant from $SU(2)$ with two fundamental flavors}
\date{\vspace{-2.5em}}

\begin{document}
\maketitle

{
\setcounter{tocdepth}{2}
\tableofcontents
}
\section{Motivation}\label{motivation}

Due to the Higgs boson's scalar nature, radiative corrections to its mass are not renormalizable unless one uses a fine-tuning procedure of parameters. However, the idea that the universe is indeed a realization of the exact necessary fine-tuned parameters seems unnatural. It stands to reason that an underlying mechanism or principle should fix the Higgs mass where we find it experimentally.

One promising proposal is the existence of an underlying strongly coupled sector of massless quarks that replaces the weak sector. In this case, the Higgs boson arises as a composite particle in the spectrum of this theory, depending on the concrete realization of the vacuum alignment in the strong symmetry breaking, either as a light resonance or a pseudo-Nambu-Goldstone boson.

The number of candidates for such an additional sector is rich; one can explore different gauge groups, flavor symmetries, representations, and embedding mechanisms into the Standard Model. We rely on phenomenological constraints to select promising candidates and provide predictions for the spectrum from the lattice as input for experimental physicists. While the strongly coupled dynamics of this additional sector guarantee asymptotic freedom and the existence of the scalar singlet in the spectrum that could act as the Higgs boson, we can constrain the parameter space by requiring the preservation of a custodial symmetry that fixes the value of the \(\rho\) parameter. A theory with a $SU(2)$ gauge symmetry and two mass degenerate fermions in the fundamental representation given by
\begin{equation}
\mathcal{L} = -\dfrac{1}{4}F^{a}_{\mu\nu}F^{a\mu\nu} + \bar{u}(\mathrm{i}\gamma^{\mu}D_{\mu} - m)u + \bar{d}(\mathrm{i}\gamma^{\mu}D_{\mu} - m)d\,,
\end{equation}
with the \(u\) and \(d\) fields acting as the additional mass-degenerate up- and down-quark analoga in the new sector in the chiral limit is the minimal scenario expressing this property and is, therefore, a promising candidate for a composite-Higgs theory that could replace the weak sector in the Standard Model while preserving its features, see \citep{Cacciapaglia:2020kgq} for a review.

In this proceedings, we aim to precisely evaluate the pseudoscalar decay constant, which is central to setting the scale for phenomenological predictions for this composite-Higgs model from lattice calculations.

\section{Simulation strategy}\label{simulation-strategy}

We aim to generate ensembles at masses that are lighter than all available studies for this theory, requiring the use of state-of-the-art lattice techniques. For this, we are using \texttt{HiRep}, a highly optimized codebase with recently added support for GPU acceleration \citep{DelDebbio:2008zf, Martins:2024dew, Martins:2024sdd}, offsetting the additional computational expense at fine lattices and Dirac inversions at low condition numbers. Apart from increasing the computational resources available, we are employing Hasenbusch acceleration to lower the condition number on ensembles at very light masses during ensemble generation, exponential clover improvement to increase stability and lower discretization effects, and a mixed action approach during measurements to achieve additional automatic \(\mathcal{O}(a)\)-improvement, improvement of chirality properties of the action and a reduction in statistical uncertainties for obtaining the renormalized results.

\subsection{Hasenbusch Acceleration}\label{hasenbusch-acceleration}

We use Hasenbusch acceleration \citep{Hasenbusch:2001ne}, with two additional levels. This acceleration relies on the fact that we can split the fermion determinant
\begin{equation}
\mathrm{det}(MM^{\dagger}) \propto \int \mathcal{D}\psi\mathcal{D}\psi^{\dagger}\mathrm{exp}(-|M^{-1}\psi|^2)
\end{equation}
for an operator \(M\) with a small condition number into more integrations
\begin{equation}
\mathrm{det}(MM^{\dagger}) \propto \int \mathcal{D}\psi\mathcal{D}\psi^{\dagger}\mathcal{D}\phi\mathcal{D}\phi^{\dagger}\mathrm{exp}(-|\tilde{M}^{-1}\psi|^2)\mathrm{exp}(-|\tilde{M}M^{-1}\phi|^2)
\end{equation}
yielding two separate operators \(\tilde{M}\) and \(M\tilde{M}^{-1}\) defined over a new bare mass and mass shift whose inversion is better conditioned. Repeating this with more levels can reduce the Hamiltonian violations at small masses at a fixed number of steps and, therefore, reduce total trajectory execution times.

\subsection{\texorpdfstring{Non-perturbative \(\mathcal{O}(a)\)-improvement in ensemble generation}{Non-perturbative \textbackslash mathcal\{O\}(a)-improvement in ensemble generation}}\label{non-perturbative-mathcaloa-improvement-in-ensemble-generation}

The previous study in \citep{Arthur:2016dir} shows that using unimproved Wilson Fermions yields results that are strongly affected by discretization effects. To reduce these effects, improving the stability of the simulations and minimizing exceptional configurations, we are employing Wilson Fermions with a non-perturbative exponential clover improvement, as proposed in \citep{Francis:2019muy}, defined by the full Dirac operator
\begin{equation}
D_{\mathrm{exp}} = \dfrac{1}{2}\sum_{\mu}\left(\gamma_\mu (\nabla_{\mu}^{\ast} + \nabla_{\mu}) - a\nabla_{\mu}^{\ast}\nabla_{\mu}\right) + (4+m_0)\,\mathrm{exp}\left(\dfrac{c_{\mathrm{sw}}}{4+m_0}\dfrac{\mathrm{i}}{4}\sigma_{\mu\nu}\hat{F}_{\mu\nu}\right)\,.
\end{equation}
\begin{figure}
\centering
\includegraphics[width=0.75\linewidth]{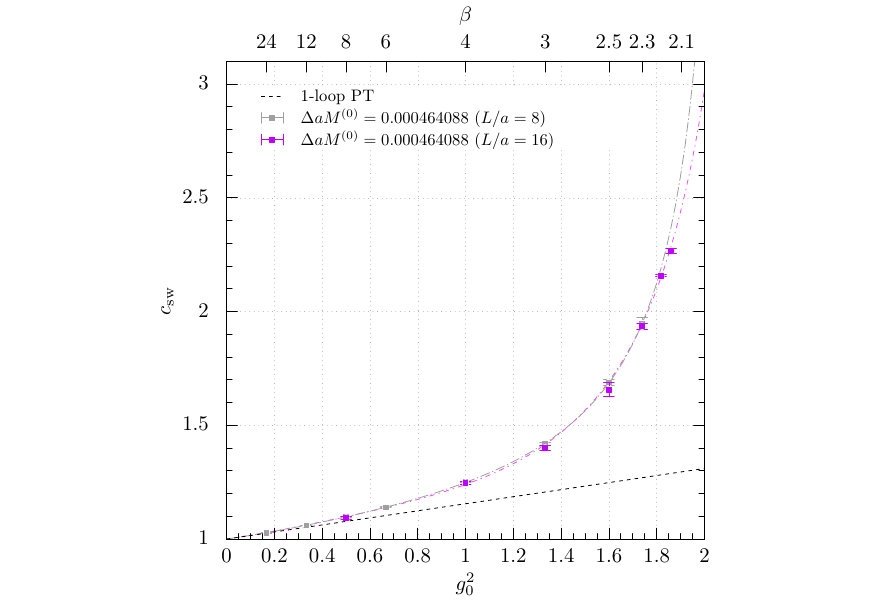}
\caption{Tuning of the $c_{\mathrm{sw}}$ parameter depending on the coupling $\beta$ to achieve the non-perturbative $\mathcal{O}(a)$ improvement, see also \cite{Bowes:2023ihh, laurence2024}.}
\label{csw_np}
\end{figure}

Here, \(c_{\mathrm{sw}}\) is tuned using Schr\"{o}dinger functional ensembles, matching the renormalized quark mass and mass shift to its tree-level values \citep{Luscher:1996ug}, Fig.~\ref{csw_np} is also published as part of the work in \citep{Bowes:2023ihh, laurence2024}. In the scaling regime, we find a smooth curve for the value of \(c_{\mathrm{sw}}\) depending on the coupling \(\beta\). This indicates that we are in a suitable regime to extrapolate to the continuum. In this work, we examine the three coarsest points in Fig.~\ref{csw_np}, which have autocorrelation times in the topological charge of 5 to 50 trajectories. This is the first study that examines the behavior in stability and \(\mathcal{O}(a^2)\) effects using and exponential clover improvement for an $SU(2)$ gauge group.

\subsection{Chirally-improved measurements and renormalization}\label{chirally-improved-measurements-and-renormalization}

While highly performant, Wilson fermions break chiral symmetry and are therefore not ideal for targeting simulations close to the chiral limit. To combine computational feasibility and good chiral properties \citep{Bar:2002nr} originally proposed to generate ensembles with a cheap action and then perform measurements with a chirally improved setup. For this, we add a chirally rotated, twisted mass \citep{Frezzotti:1999vv, Frezzotti:2000nk, Frezzotti:2001ea, Sint:2007ug} term
\begin{equation}
D_{\mathrm{exp,TM}}(m_0, \mu_0) = D_{\mathrm{exp}}(m_0) + \mathrm{i}\gamma_5\tau^{3}\mu_0\,,
\end{equation}
and tune the parameters \(m_0\) and \(\mu_0\) to maximal twist, i.e.~tuning the PCAC mass to zero and the pseudoscalar mass in the valence sector to the pseudoscalar mass in the sea. In addition to the improved chiral properties, this yields automatic \(\mathcal{O}(a)\) improvement through the symmetries of the action. This improvement is in addition to the exponential clover improvement.

Another central advantage is the property that the pseudoscalar decay constant does not need renormalization at maximal twist, see \citep{Shindler:2007vp}, avoiding a source of error and allowing a more precise scale setting of the theory as previously achieved in \citep{Hernandez:2019qed}. This is because the pseudoscalar decay constant defined as
\begin{equation}
f_{\mathrm{PS}}^{\mathrm{R}} = \dfrac{2M^{\mathrm{R}}\langle 0 | P | \pi\rangle}{M_{\mathrm{PS}}^2} = Z_{A}f_{\mathrm{PS}}
\end{equation}
depends on the renormalized quark mass $M^{\mathrm{R}}$, which, for Wilson fermions without a twisted mass term, is given by the PCAC mass, which needs to be renormalized multiplicatively. However, when including the twisted mass term, the renormalized quark mass is  $M^{\mathrm{R}} = \sqrt{(Z_A m_{\mathrm{PCAC}})^2 + (a\mu_0)^2}$. 
As a result the tuning to maximal twist at $m_{\mathrm{PCAC}}=0$ implies $M^{\mathrm{R}} = \mu_0$. This simplifies the formula, as no renormalization constants appear in the formula
\begin{equation}
f_{\mathrm{PS}}^{\mathrm{R}} = \dfrac{2\mu_0\langle 0 | P | \pi\rangle}{M_{\mathrm{PS}}^2}
\end{equation}
which only depends on the bare twisted mass \(\mu_0\), the pseudoscalar mass \(M_{\mathrm{PS}}\) and the matrix element \(\langle 0 | P | \pi\rangle\). As \citep{Hernandez:2019qed} points out, any inexact tuning can be corrected using the formula
\begin{equation}
f_{\mathrm{PS}}^{\mathrm{R}} = f_{\mathrm{PS}}\sqrt{1 + \left(\dfrac{Z_Am_{\mathrm{PCAC}}^{\mathrm{R}}}{a\mu_0}\right)}
\end{equation}
requiring the renormalization constant \(Z_A\).

\section{Ensemble generation}\label{ensemble-generation}

\begin{table}
\centering
\caption{Overview of ensemble parameters}
\begin{tabular}{cS[table-format=1.2]S[table-format=1.4]S[table-format=1.7]S[table-format=1.8]r}
\toprule
Lattice & $\beta$ & $m$ & $w_{0}/a$ & $aM_{\mathrm{PS}}$ & $af_{\mathrm{PS}}$ \\
\midrule
$48\times24^3$ & 2.15 & -0.2645 & 3.028(11) & 0.3390(14) & 0.05051(14) \\
$48\times24^3$ & 2.15 & -0.2624 & 2.962(9)  & 0.3572(18) & 0.05220(19) \\
$48\times24^3$ & 2.2  & -0.269  & 3.558(18) & 0.2804(13) & 0.04154(19) \\
$48\times24^3$ & 2.2  & -0.2657 & 3.498(15) & 0.3036(16) & 0.04348(17) \\
$48\times24^3$ & 2.2  & -0.26   & 3.341(14) & 0.3377(16) & 0.04695(21) \\
$36^4$ &         2.3  & -0.29   & 5.201(39) & 0.2004(9) & 0.02827(14) \\
\bottomrule
\end{tabular}
\label{tbl-ensembles}
\end{table}

Generated ensembles are available in Tbl.~\ref{tbl-ensembles}. Here, we generated one ensemble at \(\beta=2.3\), the finest ensemble for which the decorrelated statistics are sufficient to this date, and two more ensembles at \(\beta=2.2\) and \(\beta=2.15\). The results of these ensembles will be interpolated to the pseudoscalar mass of the reference ensemble to take the mass-dependent continuum limit. A chiral extrapolation will then rely on taking this mass to zero, which will be done in the future.

\begin{figure}
\centering
\includegraphics[width=12cm]{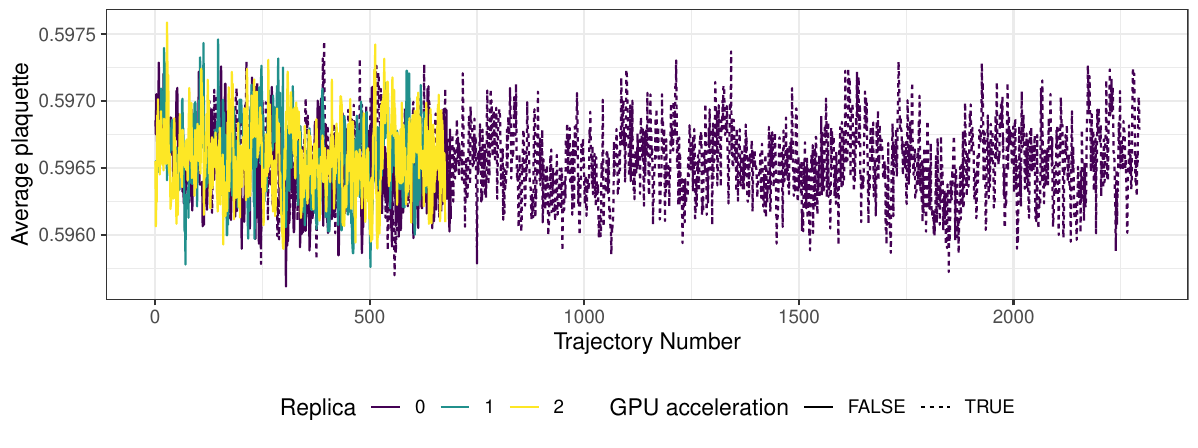}
\caption{Comparison of the average plaquette of the CPU and GPU ensembles}
\end{figure}

The ensembles used in \citep{Drach:2021uhl} were not chiral enough to reach the regime where we can fit the prediction from chiral perturbation theory. We aim to generate ensembles at lighter quark mass than in this paper. To control the chirality of our ensembles, we use the ratio \(\tfrac{m_{\mathrm{V}}}{m_{\mathrm{PS}}}\), which is 1 in the limit of infinite quark mass and goes to infinity towards the chiral limit. To reach these low quark masses on fine lattice spacings with sufficient \(M_{\mathrm{PS}}L\), we need to rely on GPU acceleration. One ensemble presented in this work was generated on LUMI-G AMD MI250X GPUs. We found that, depending on the concrete lattice size and parameters, a single GPU node replaces 10-14 CPU nodes on DIaL3 (DiRAC Data Intensive service at the University of Leicester) with 128 AMD EPYC 7742 cores. More benchmarks and tests can be found in \citep{Martins:2024dew, Martins:2024sdd}.

\section{Continuum Extrapolation}\label{continuum-extrapolation}

We now interpolate the values for the pseudoscalar decay constant and the lattice spacing in units of \(w_0\) for the ensembles with \(\beta=2.2\) and \(\beta=2.15\) to the pseudoscalar mass of the reference ensemble at \(\beta=2.3\). We expect finite-volume effects to be negligible, since \(M_{\mathrm{PS}}L\) is between \(6.7\) and \(8.4\) and for the determination of \(w_0\) we are using a smearing radius of \(r \leq 0.4L\).

\begin{figure}
\centering
\begin{minipage}[t]{0.461\textwidth}
\centering
\includegraphics[trim={0 0 2cm 0},clip,width=\linewidth]{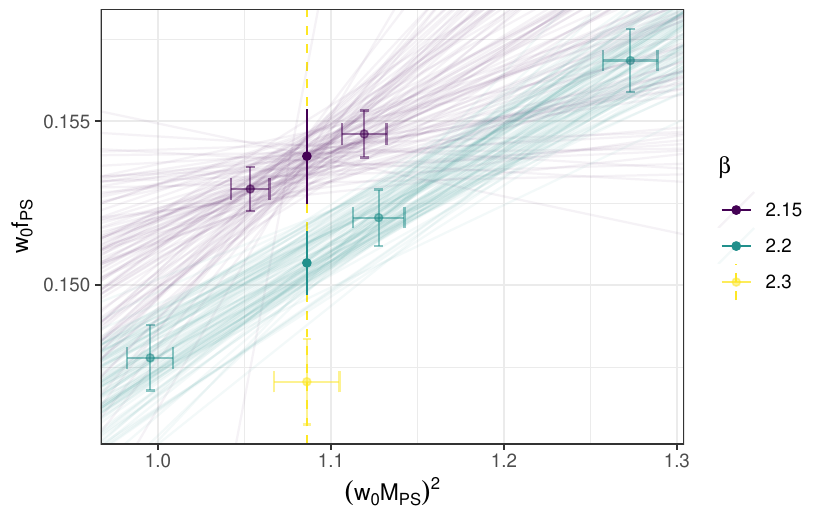}
\end{minipage}%
\begin{minipage}[t]{0.539\textwidth}
\includegraphics[width=\linewidth]{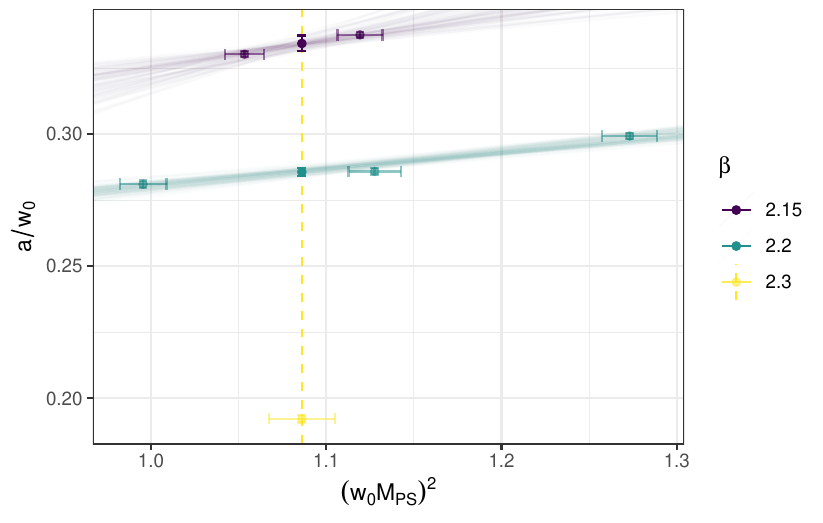}
\end{minipage}
\caption{Interpolation to reference mass of the pseudoscalar decay constant (left) and lattice spacing (right) in units of $w_0$}
\end{figure}

\begin{figure}
\centering
\includegraphics[width=8cm]{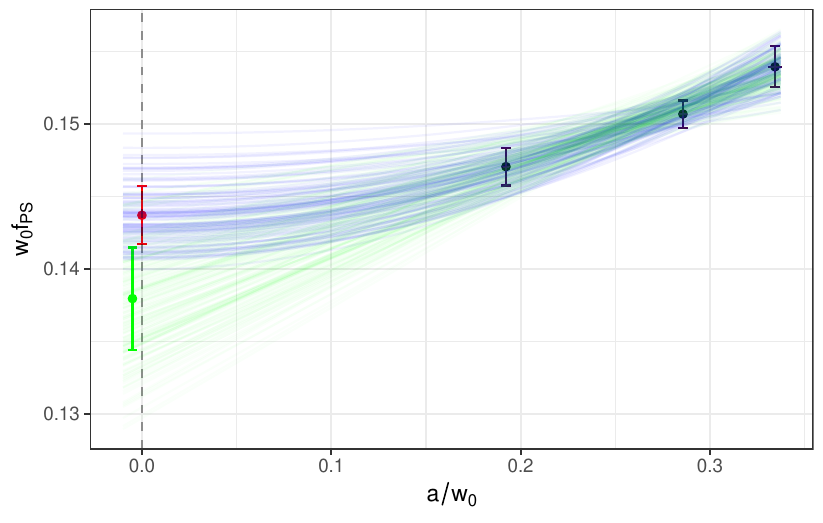}
\caption{Continuum extrapolation of \(w_0f_{\mathrm{PS}}\)}
\end{figure}

Both the ensemble generation with the exponential clover improved action and the twisted mass measurements feature an \(\mathcal{O}(a)\) improvement, so, if the improvement is exact enough, we should see an \(\mathcal{O}(a^2)\) behavior in the continuum limit. The behavior of our points resembles this behavior. However, we need one more point with finer lattice spacing to confirm that the $\mathcal{O}(a)$ effects are eliminated. Nevertheless, the result from a linear fit is only in slight tension with the actual continuum result at \(w_0f_{\mathrm{PS}} = 0.1436(19)\) that we find for a reference mass of \((w_0M_{\mathrm{PS}})^2 = 1.09(2)\).

This is because the discretization effects are very small and in the region of less than \(10\%\) for the given lattice spacings. The study previously published in \citep{Arthur:2016dir} using a non-perturbatively renormalized evaluation of the pseudoscalar decay constant using unimproved Wilson fermions showed discretization effects in a comparable region of around \(30\%\).

\section{Conclusion and outlook}\label{conclusion-and-outlook}

We took the continuum limit for the pseudoscalar decay constant at a fixed reference mass at \(\tfrac{m_{\mathrm{V}}}{m_{\mathrm{PS}}}\approx 1.5\) using exponential clover improvement in ensemble generation and a twisted mass evaluation of \(f_{\mathrm{PS}}\) at maximal twist and achieved high precision with small \(\mathcal{O}(a^2)\) effects. This is the first study showing the size of discretization effects for exponential clover improvement for $SU(2)$.

However, we need more data towards the continuum limit to be sure that the \(\mathcal{O}(a)\) effects are truly eliminated, and we need many more chiral points to extrapolate to the chiral limit. GPU acceleration is central for these simulations since configuration generation times are already high. More ensembles closer to the chiral limit and a full scattering analysis will be completed in the future.

\section{Acknowledgements}\label{acknowledgements}

This project has received funding from the European Union's Horizon 2020 research and innovation program under the Marie Sk\l odowska-Curie grant agreement \textnumero 813942 and supported with resources on LUMI-G provided by the Danish eInfrastructure Consortium under grant applications DeiC-SDU-N5-2024055 and DeiC-SDU-L5-16. The work of FRL is supported in part by Simons Foundation grant 994314 (Simons Collaboration on Confinement and QCD Strings) and the Mauricio and Carlota Botton Fellowship. 

This work used the DiRAC Data Intensive service (DIaL2 \& DIaL) at the University of Leicester, managed by the University of Leicester Research Computing Service on behalf of the STFC DiRAC HPC Facility (www.dirac.ac.uk). We are also grateful for access to the DiRAC Extreme Scaling service Tursa at the University of Edinburgh, managed by the Edinburgh Parallel Computing Centre on behalf of the STFC DiRAC HPC Facility (www.dirac.ac.uk). The DiRAC services at Edinburgh and Leicester were funded by BEIS, UKRI and STFC capital funding and STFC operations grants. DiRAC is part of the UKRI Digital Research Infrastructure. 

Part of this work was carried out using the computational facilities of the High Performance Computing Centre, University of Plymouth.

\renewcommand\refname{References}
  \bibliography{literature.bib}

\begin{thebibliography}{10}

\bibitem{Cacciapaglia:2020kgq}
Giacomo Cacciapaglia, Claudio Pica, and Francesco Sannino.
\newblock {Fundamental Composite Dynamics: A Review}.
\newblock {\em Phys. Rept.}, 877:1--70, 2020.

\bibitem{DelDebbio:2008zf}
Luigi Del~Debbio, Agostino Patella, and Claudio Pica.
\newblock {Higher representations on the lattice: Numerical simulations. SU(2)
  with adjoint fermions}.
\newblock {\em Phys. Rev. D}, 81:094503, 2010.

\bibitem{Martins:2024dew}
Sofie Martins, Erik Kjellgren, Emiliano Molinaro, Claudio Pica, and Antonio
  Rago.
\newblock {GPU-accelerated Higher Representations of Wilson Fermions with
  HiRep}.
\newblock {\em PoS}, EuroPLEx2023:035, 2024.

\bibitem{Martins:2024sdd}
Sofie Martins, Erik Kjellgren, Emiliano Molinaro, Claudio Pica, and Antonio
  Rago.
\newblock {Scaling SU(2) to 1000 GPUs using HiRep}.
\newblock In {\em {41st International Symposium on Lattice Field Theory}}, 11
  2024.

\bibitem{Hasenbusch:2001ne}
Martin Hasenbusch.
\newblock {Speeding up the hybrid Monte Carlo algorithm for dynamical
  fermions}.
\newblock {\em Phys. Lett. B}, 519:177--182, 2001.

\bibitem{Arthur:2016dir}
Rudy Arthur, Vincent Drach, Martin Hansen, Ari Hietanen, Claudio Pica, and
  Francesco Sannino.
\newblock {SU(2) gauge theory with two fundamental flavors: A minimal template
  for model building}.
\newblock {\em Phys. Rev. D}, 94(9):094507, 2016.

\bibitem{Francis:2019muy}
Anthony Francis, Patrick Fritzsch, Martin L\"uscher, and Antonio Rago.
\newblock {Master-field simulations of O($a$)-improved lattice QCD: Algorithms,
  stability and exactness}.
\newblock {\em Comput. Phys. Commun.}, 255:107355, 2020.

\bibitem{Bowes:2023ihh}
Laurence~Sebastian Bowes, Vincent Drach, Patrick Fritzsch, Antonio Rago, and
  Fernando Romero-Lopez.
\newblock {2-flavour $SU(2)$ gauge theory with exponential clover Wilson
  fermions}.
\newblock {\em PoS}, LATTICE2023:094, 2024.

\bibitem{laurence2024}
Laurence~Sebastian Bowes, Vincent Drach, Patrick Fritzsch, Sofie Martins,
  Antonio Rago, and Fernando Romero-López.
\newblock {The mass of the $\sigma$ in a chiral ensemble in $SU(2)$ with two
  fundamental flavours}.
\newblock 2024.
\newblock LATTICE 2024, to appear.

\bibitem{Luscher:1996ug}
Martin Luscher, Stefan Sint, Rainer Sommer, Peter Weisz, and Ulli Wolff.
\newblock {Nonperturbative O(a) improvement of lattice QCD}.
\newblock {\em Nucl. Phys. B}, 491:323--343, 1997.

\bibitem{Bar:2002nr}
Oliver Bar, Gautam Rupak, and Noam Shoresh.
\newblock {Simulations with different lattice Dirac operators for valence and
  sea quarks}.
\newblock {\em Phys. Rev. D}, 67:114505, 2003.

\bibitem{Frezzotti:1999vv}
Roberto Frezzotti, Pietro~Antonio Grassi, Stefan Sint, and Peter Weisz.
\newblock {A Local formulation of lattice QCD without unphysical fermion zero
  modes}.
\newblock {\em Nucl. Phys. B Proc. Suppl.}, 83:941--946, 2000.

\bibitem{Frezzotti:2000nk}
Roberto Frezzotti, Pietro~Antonio Grassi, Stefan Sint, and Peter Weisz.
\newblock {Lattice QCD with a chirally twisted mass term}.
\newblock {\em JHEP}, 08:058, 2001.

\bibitem{Frezzotti:2001ea}
Roberto Frezzotti, Stefan Sint, and Peter Weisz.
\newblock {O(a) improved twisted mass lattice QCD}.
\newblock {\em JHEP}, 07:048, 2001.

\bibitem{Sint:2007ug}
Stefan Sint.
\newblock {Lattice QCD with a chiral twist}.
\newblock In {\em {Workshop on Perspectives in Lattice QCD}}, 2 2007.

\bibitem{Shindler:2007vp}
Andrea Shindler.
\newblock {Twisted mass lattice QCD}.
\newblock {\em Phys. Rept.}, 461:37--110, 2008.

\bibitem{Hernandez:2019qed}
Pilar Hern\'andez, Carlos Pena, and Fernando Romero-L\'opez.
\newblock {Large $N_c$ scaling of meson masses and decay constants}.
\newblock {\em Eur. Phys. J. C}, 79(10):865, 2019.

\bibitem{Drach:2021uhl}
Vincent Drach, Patrick Fritzsch, Antonio Rago, and Fernando Romero-L\'opez.
\newblock {Singlet channel scattering in a composite Higgs model on the
  lattice}.
\newblock {\em Eur. Phys. J. C}, 82(1):47, 2022.

\end{thebibliography}

\end{document}